\documentclass[pre,aps,byrevtex,showpacs,amsmath,amssymb,preprint]{revtex4}
\usepackage{times}
\usepackage{epsf}
\usepackage{graphicx}
\usepackage{dcolumn}
\usepackage{bm}

\begin{document}

\title{Geometrically controllable electric fields}

\date{\today}

\author{C. Z. Fan\footnote{Electronic address: 052019029@fudan.edu.cn}, Y. Gao, and J. P.
Huang\footnote{Corresponding author. Electronic address:
jphuang@fudan.edu.cn}}

 \affiliation{Surface Physics Laboratory (National key
laboratory) and Department of Physics, Fudan University, Shanghai
200433, China}

\begin{abstract}

According to a first-principles approach, we clarify electric
cloaks' universality concerning  a new class of phase transitions
between negative pathway (NGP) and normal pathway of electric
displacement fields, which are driven by the geometric shape of
the cloak. We report that  the NGP arises from shape-enhanced
strong negative electric polarization, and that it is related to a
symmetric oscillation of the paired  electric permittivities,
which are shown to satisfy a sum rule. The NGP does not occur for
a spherical cloak, but appears up to maximum as the ratio $a/b$
between the long- and short- principal axis of the spheroidal
cloak is about 5/2, and eventually disappears as $a/b$ becomes
large enough corresponding to a rod-like shape. Then, the cloaking
efficiency is   compared between different geometrical shapes. The
possibility of experiments is discussed. This work has relevance
to crucial control of electric fields and to general physics of
phase transitions.

\end{abstract}



\maketitle

\clearpage
\newpage

1 Introduction

Metamaterials~\cite{VeselagoSP68,PendryPRL96} are a
new class of electromagnetic materials, which are currently under
extensive
investigation~\cite{SmithS04,PendryS06,SchurigOE06,SchurigS06,RuanPRL07,YanPRL07,ChenPRL07,CaiNP07}.
They owe their properties to subwavelength details of structure
rather than to their chemical composition, and can be designed to
have properties difficult or impossible to find in nature. An
electromagnetic spherical cloak is well-known for hiding objects
from electro-magnetic
waves~\cite{PendryS06,SchurigOE06,SchurigS06,CaiNP07}, due to the
high freedom of design of metamaterials~\cite{SmithS04}, which may
be both inhomogeneous and anisotropic in their electric permittivity
and magnetic
permeability~\cite{SmithS04,PendryS06,SchurigOE06,SchurigS06,RuanPRL07,YanPRL07,ChenPRL07}.
Owing to intriguing potential applications, it has received
extensive attention, e.g., ranging from its  two-dimensional
counterpart~\cite{SchurigOE06,RuanPRL07,YanPRL07} and scattering
cross section~\cite{ChenPRL07}, to its extensions like acoustic
cloaks~\cite{ChenAPL07}. In this work, by using a first-principles
approach based on the coordinate transformation method, we
investigate electric cloaks and clarify their universality
concerning a new class of  phase transition between negative pathway
(NGP) and normal pathway (NMP) of electric displacement fields as
geometric shape of the cloak changes in a typical range. The
shape-driven phase transition seems to be of particular interest
when compared with abundant phase transitions already known in
condensed matter and nature. We shall also calculate cloaking
efficiency of the electric cloaks with various shapes. This work has
relevance to crucial control of electric fields and to general
physics of phase transitions.

The remainder of this work is organized as follows. In Section 2,
based on the coordinate transformation method, we present a
first-principles approach to derive the expressions for electric
permittivities in electric cloaks with non-spherical shapes. This is
followed by Section 3, in which we numerically clarify their
universality concerning the phase transition between negative
pathway and normal pathway of electric displacement fields, and
calculate cloaking efficiency under various conditions. This paper
ends with a discussion and conclusion in Section 4.

2 Formalism

To proceed, we shall  use a first-principles approach, namely, the
coordinate transformation method~\cite{PendryS06,SchurigOE06}. We
consider that each point in the original Cartesian coordinates can
be represented as $(x, y, z)$, while in the distorted coordinates it
can be given as $(u, v, w)$, which is the location of the new point
with respect to the $x$, $y$, and $z$ axes, namely, $u = u(x, y,
z)$, $v = v(x, y, z)$, and $w = w(x, y, z)$~\cite{PendryS06}. In the
distorted coordinates, the form of Maxwell's Equations are
invariant, but the corresponding electric permittivity are got in
the following form $\epsilon_u^\prime=\epsilon_u
\frac{Q_uQ_vQ_w}{Q_u^2}$ ($\epsilon_u$: electric permittivity in the
original Cartesian coordinates),
where
\begin{eqnarray}
Q_u^2&=&\left(\frac{\partial x}{\partial
u}\right)^2+\left(\frac{\partial y}{\partial
u}\right)^2+\left(\frac{\partial z}{\partial u}\right)^2,\\
Q_v^2&=&\left(\frac{\partial x}{\partial
v}\right)^2+\left(\frac{\partial y}{\partial
v}\right)^2+\left(\frac{\partial z}{\partial v}\right)^2,\\
Q_w^2&=&\left(\frac{\partial x}{\partial
w}\right)^2+\left(\frac{\partial y}{\partial
w}\right)^2+\left(\frac{\partial z}{\partial w}\right)^2.
\end{eqnarray}
Now we consider an ellipsoidal cloak in three dimensions. In the
Cartesian coordinates $(x, y, z)$, the equation for describing an
ellipsoidal shape is
\begin{eqnarray}
\frac{x^2}{a^2}+\frac{y^2}{b^2}+\frac{z^2}{c^2}=1,\nonumber
\end{eqnarray}
where $a$, $b$, and $c$ are the three principal semi-axes of the
ellipsoid, respectively. Next, we choose $(\lambda, \mu, \nu)$ to
represent a point in the ellipsoidal coordinates. So far, the
relation between $(\lambda, \mu, \nu)$ and $(x, y, z)$ is given as
\begin{eqnarray}
x^2&=&\frac{\left(a^2+\lambda \right) \left(a^2+\mu \right)
\left(a^2+\nu
   \right)}{\left(a^2-b^2\right) \left(a^2-c^2\right)},\\
y^2&=&\frac{\left(b^2+\lambda \right) \left(b^2+\mu \right)
\left(b^2+\nu
   \right)}{\left(b^2-c^2\right) \left(b^2-a^2\right)},\\
z^2&=&\frac{\left(c^2+\lambda \right) \left(c^2+\mu \right)
\left(c^2+\nu
   \right)}{\left(c^2-a^2\right) \left(c^2-b^2\right)}.
\end{eqnarray}
Then, we squeeze the ellipsoidal volume into an ellipsoidal shell
through the following relations
\begin{eqnarray}
 \lambda'&=& a_1+\lambda\frac{a_2-a_1}{a_2},\\
\mu'&=& b_1+\mu\frac{b_2-b_1}{b_2},\\
\nu'&=& c_1+\nu\frac{c_2-c_1}{c_2},
\end{eqnarray}
where $a_1$, $b_1$, and $c_1$ ($a_2$, $b_2$, and $c_2$) are the
inner (outer) three principal semi-axes of the ellipsoidal shell.
In the distorted ellipsoidal coordinates, we have the renormalized
values of electric permittivities, $\epsilon_{\lambda'}'$,
$\epsilon_{\mu'}'$, and $\epsilon_{\nu'}'$, inside the ellisoidal
cloak (shell):
\begin{eqnarray}
\epsilon_{\lambda'}'&=& \frac{a_2}{a_2-a_1}\frac{b_2}{b_2-b_1}\frac{c_2}{c_2-c_1}\frac{(\lambda'-a_1)^2}{\lambda'^2}\epsilon_{host},\label{10}\\
\epsilon_{\mu'}'&=& \frac{a_2}{a_2-a_1}\frac{b_2}{b_2-b_1}\frac{c_2}{c_2-c_1}\frac{(\mu'-b_1)^2}{\mu'^2}\epsilon_{host},\\
\epsilon_{\nu'}'&=&
\frac{a_2}{a_2-a_1}\frac{b_2}{b_2-b_1}\frac{c_2}{c_2-c_1}\frac{(\nu'-c_1)^2}{\nu'^2}\epsilon_{host},\label{12}
\end{eqnarray}
where $\epsilon_{host}$ denotes the electric permittivity outside
the cloak. From equations~(\ref{10})-(\ref{12}), we can see that
the electric permittivities are both materially anisotropic and
spatially
 inhomegeneous in the distorted ellipsoidal
coordinates. For achieving them, metamaterials can
help~\cite{SmithS04}.





3 Numerical results

3.1 Phase transition between negative pathway and normal pathway of
electric displacement fields

\begin{figure}
\includegraphics[width=350pt]{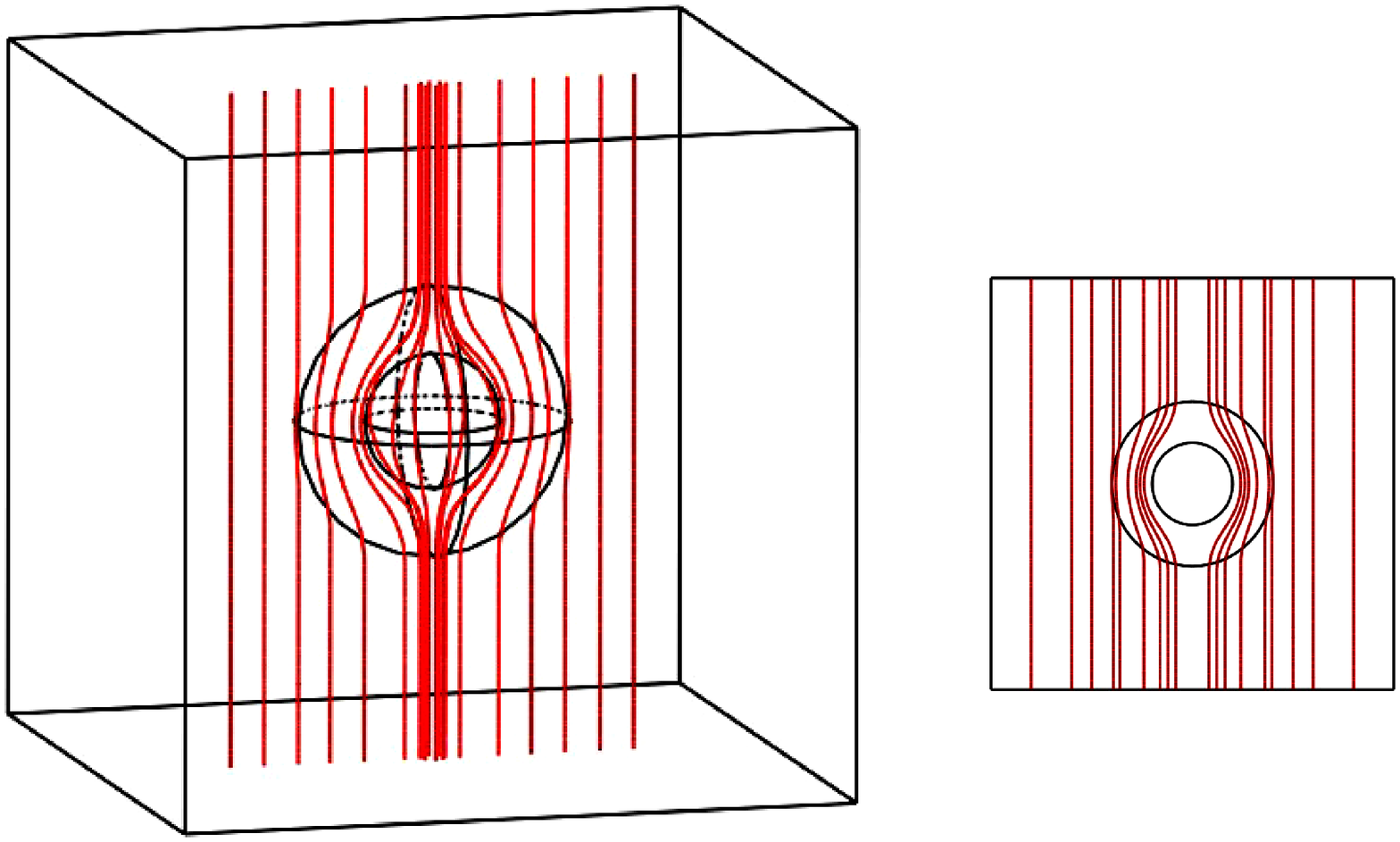}
\caption{}\label{fig1}
\end{figure}

\begin{figure}
\includegraphics[width=350pt]{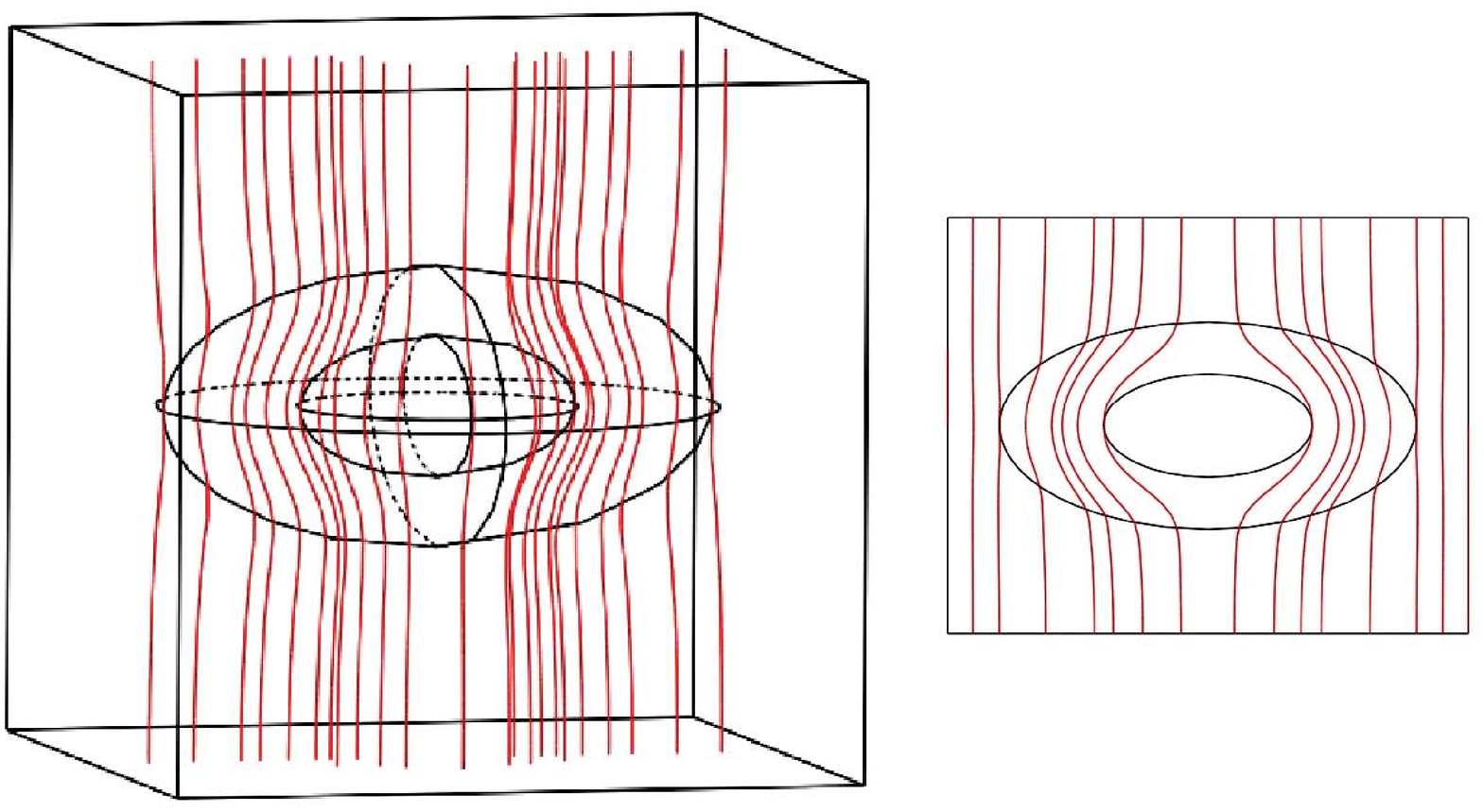}
\caption{}\label{fig2}
\end{figure}

\begin{figure}
\includegraphics[width=350pt]{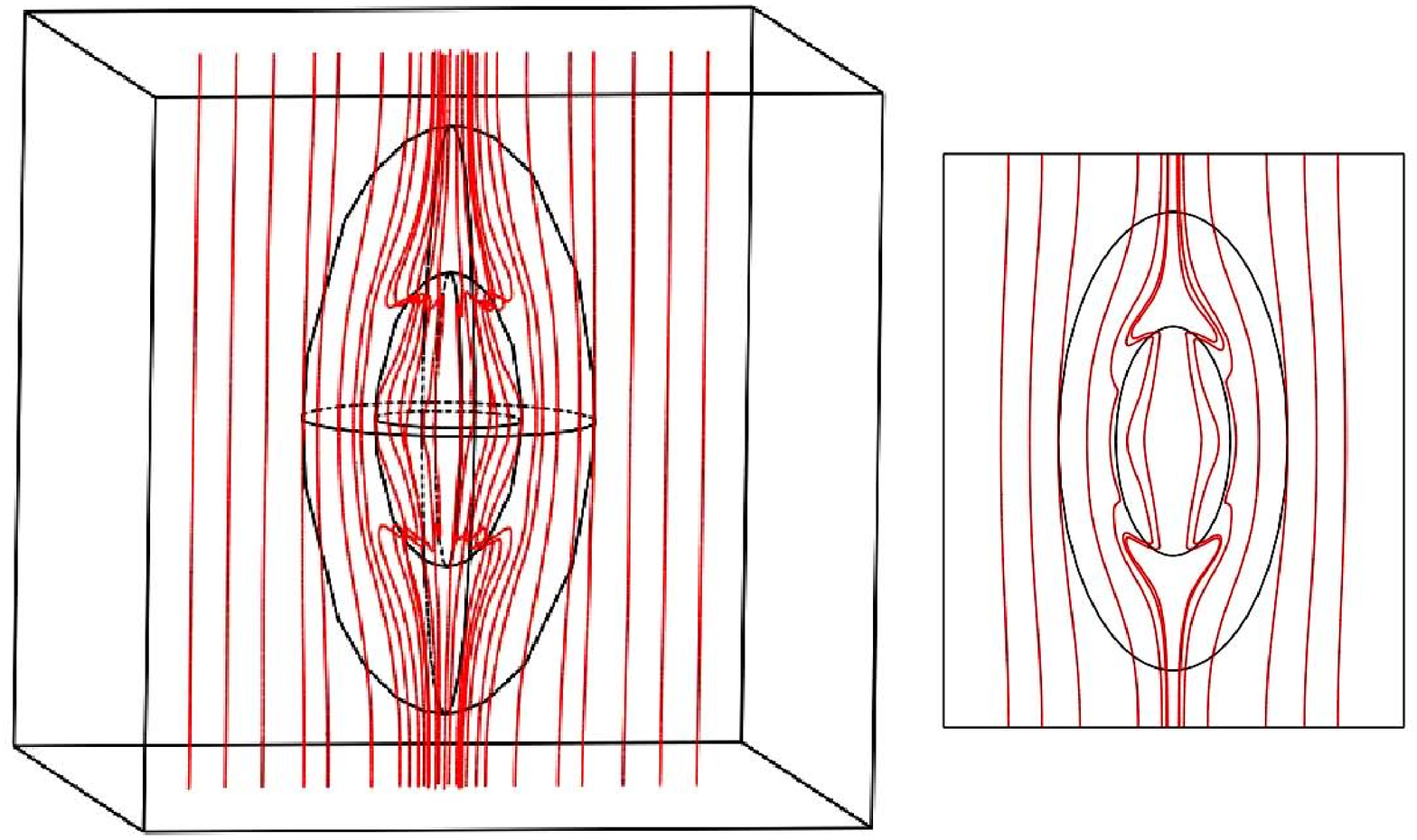}
\caption{}\label{fig3}
\end{figure}

\begin{figure}
\includegraphics[width=300pt]{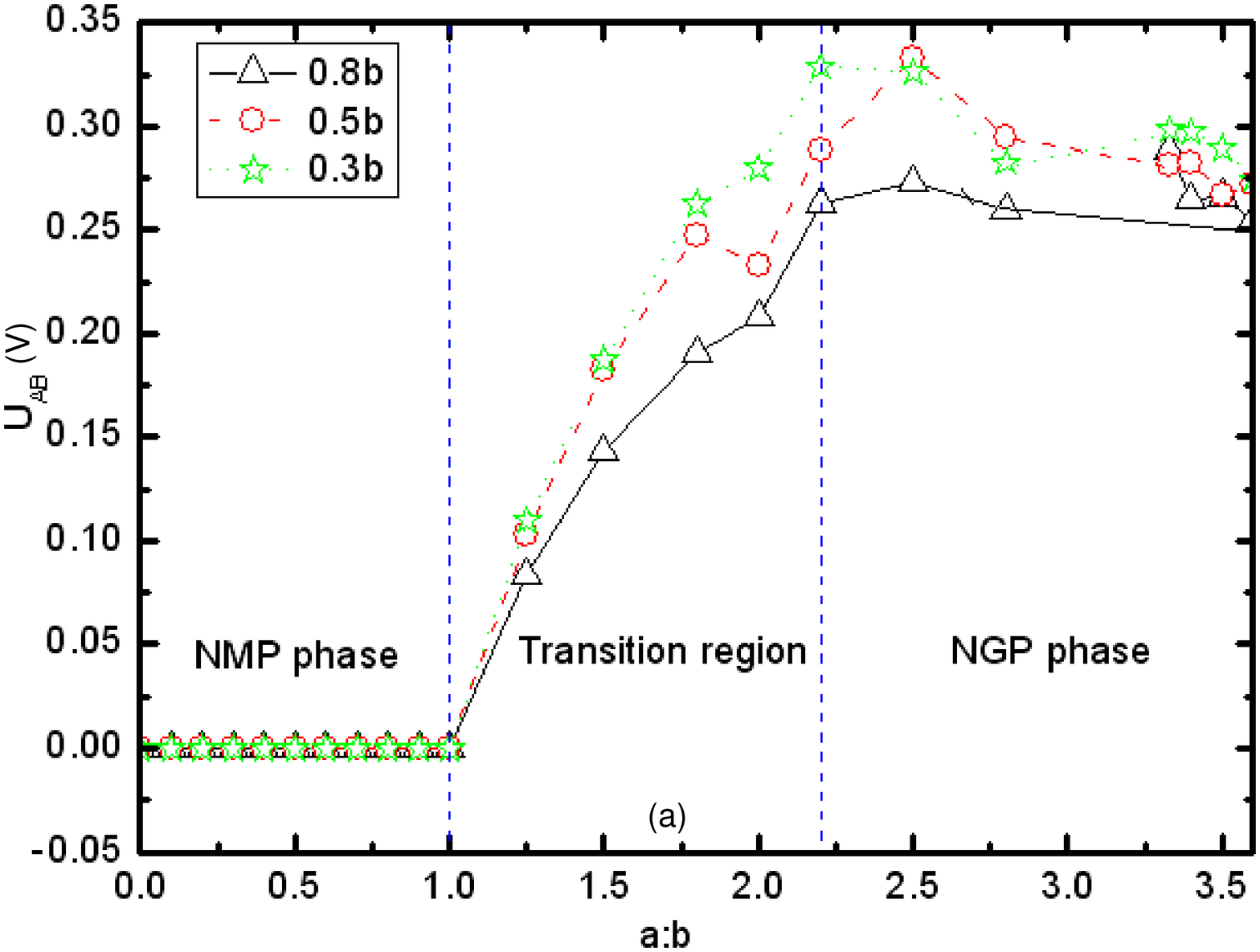}
\includegraphics[width=300pt]{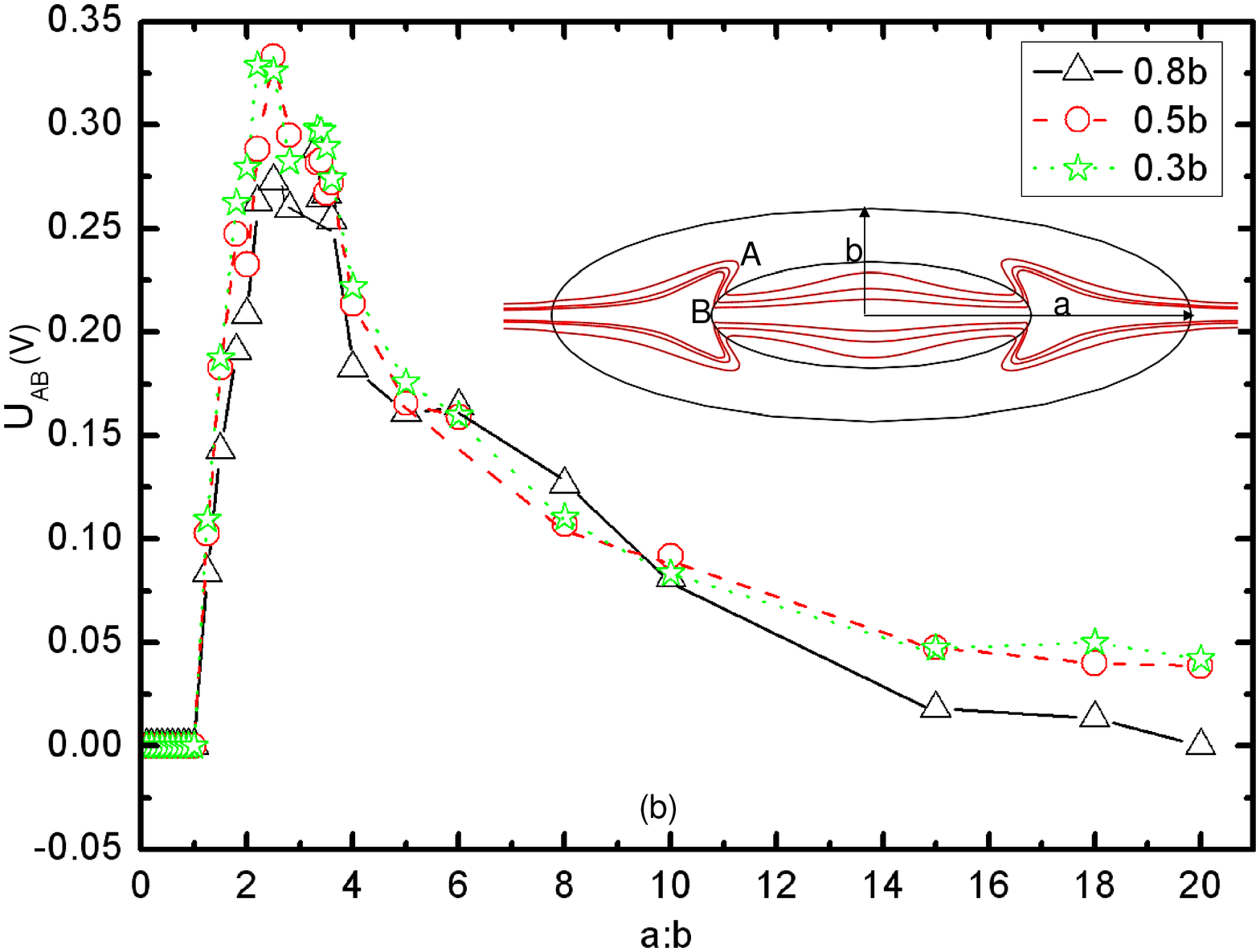}
\caption{}\label{fig5}
\end{figure}

\begin{figure}
\includegraphics[width=300pt]{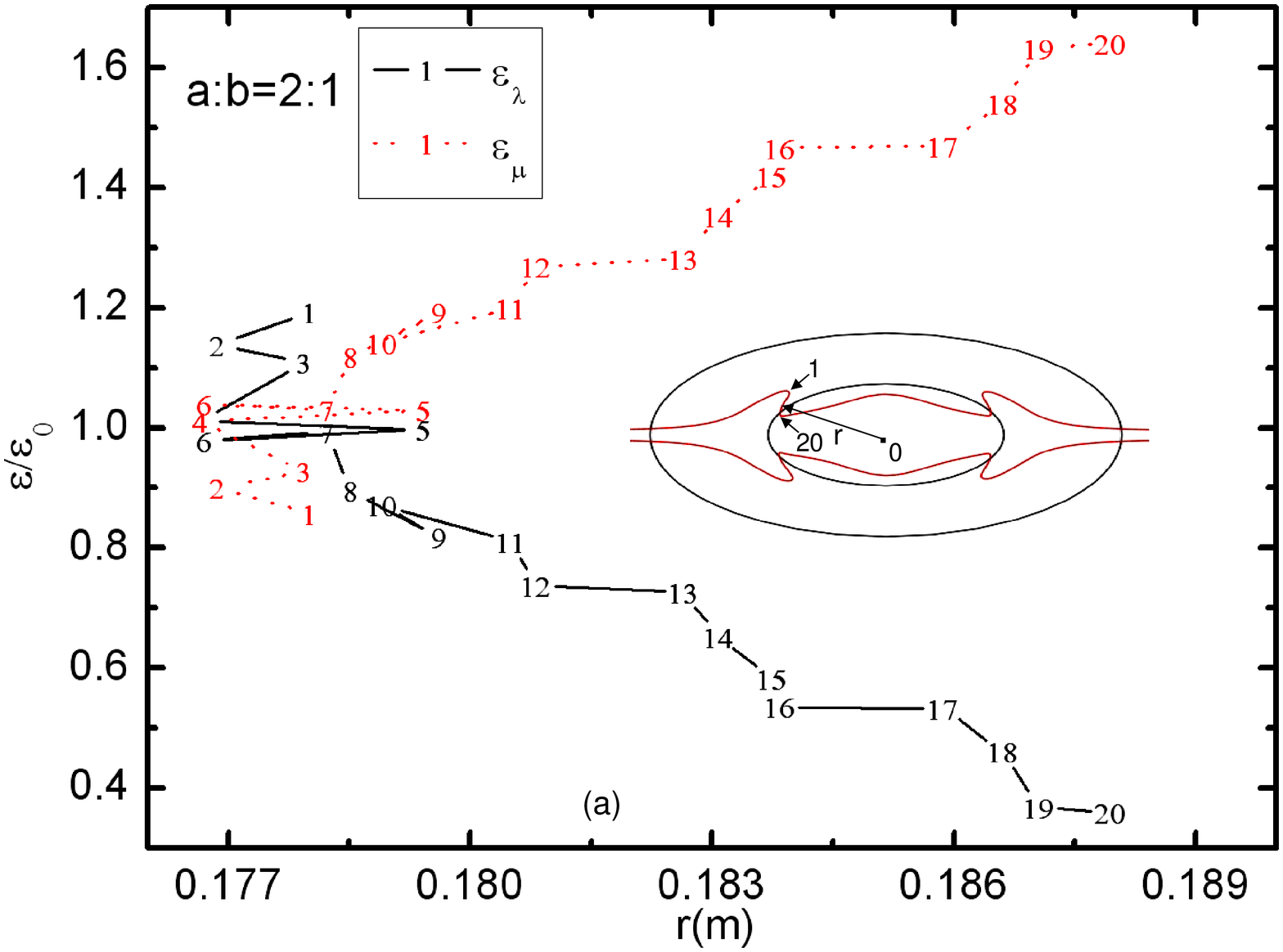}
\includegraphics[width=300pt]{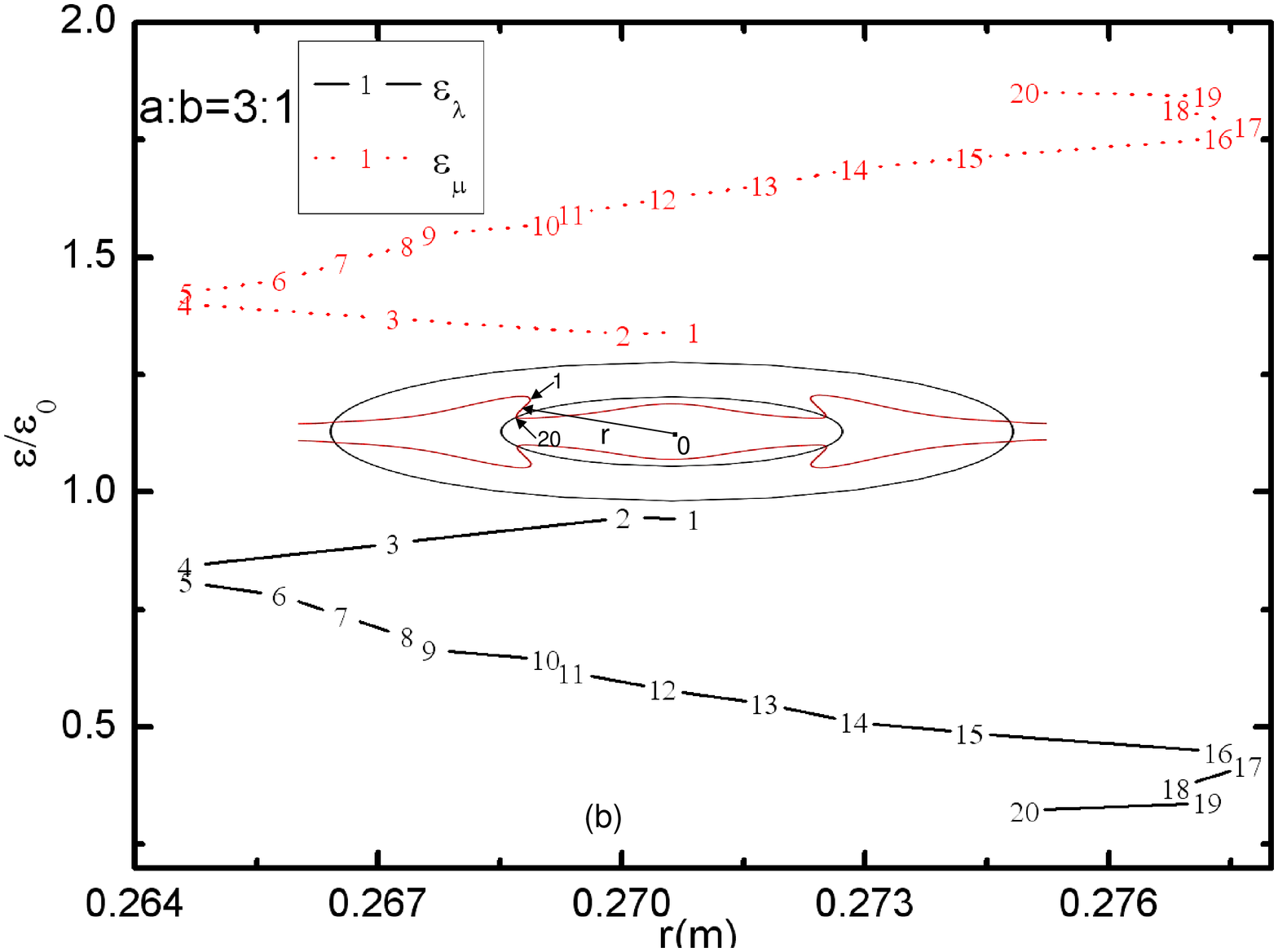}
\includegraphics[width=300pt]{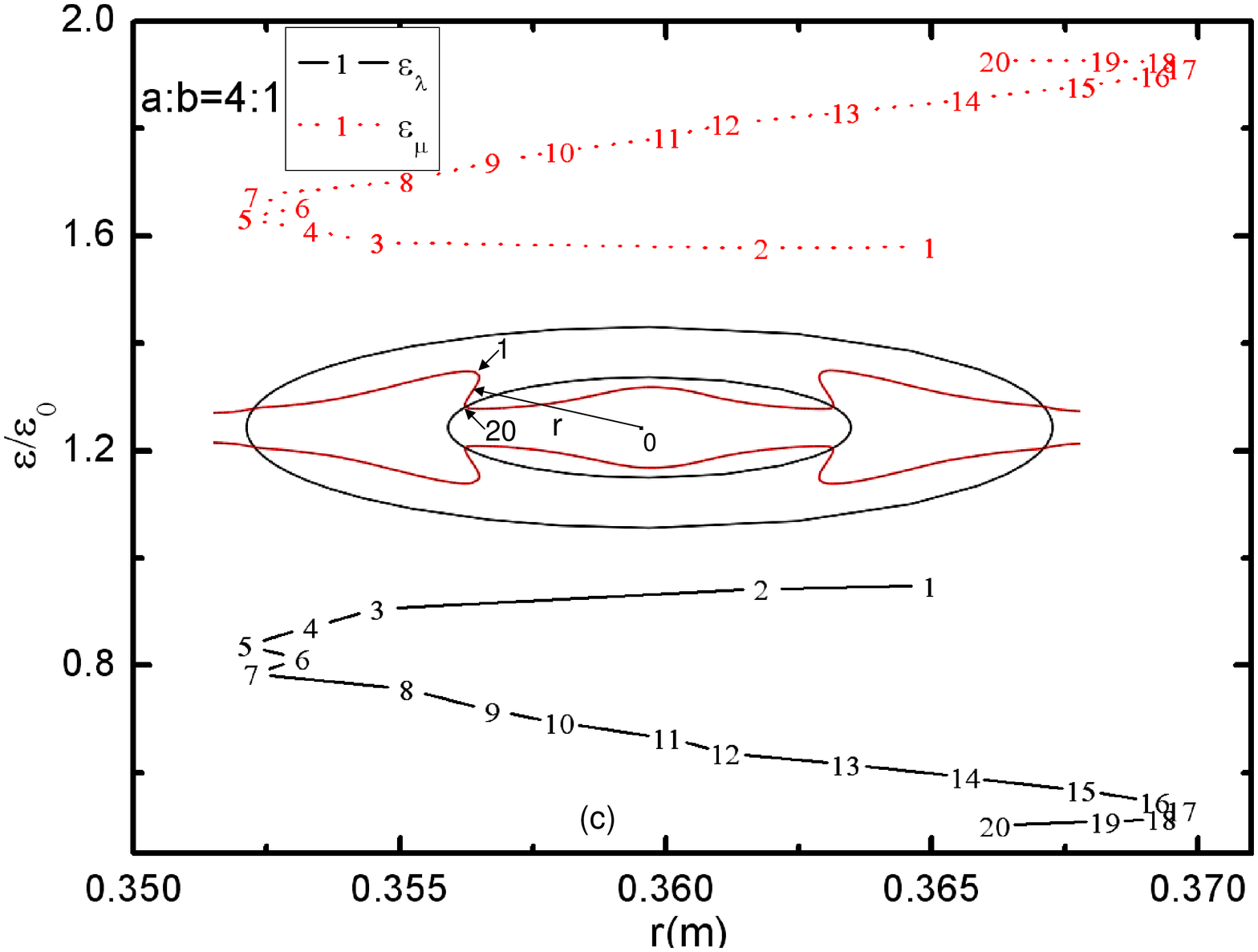}
\caption{}\label{fig6}
\end{figure}

In our numerical simulations, there are three domains in the whole
system, namely, the inner domain (domain I), the cloaking or shell
domain (domain II), and the outer domain (domain III). For
convenience, we set the electric permittivity of domain I to have
the same value as that $\epsilon_{host}$ of  domain III, and further
take $\epsilon_{host}$ to be equal to the electric permittivity of
free space $\epsilon_0$. This will not affect our results at all
because the main interest of this work is on geometrical control of
electric fields within domain II. The electric permittivity of
domain II is determined according to
equations~(\ref{10})-(\ref{12}). For all the spherical or
non-spherical cloaks discussed in Figs.~\ref{fig1}-\ref{fig6}, we
keep the volume ratio between domain II and domain I at 7:1, and the
applied electric potentials are set to be +3\,V and -3\,V at the two
opposite planes of the cubic box, respectively.
It is worth mentioning that our results  are independent of the
actual value of the potentials. As the spherical cloak is
centro-symmetric, it makes no difference when the electric field
is applied in various directions. However, for an ellipsoidal
cloak, we have to discriminate the direction of applied electric
fields due to the existence of geometric anisotropy. In our
simulations, we apply the electric potentials in two opposite
planes along the three principal axes, $a$, $b$, and $c$, of the
ellipsoidal cloak, respectively. We  focus on two types of
rotational ellipsoid (namely, spheroid) with  $a<b=c$ (oblate
spheroid) and $a>b=c$ (prolate spheroid). Throughout this work,
the set of $a, b,$ and $c$ will be used to denote both three inner
principal semi-axes ($a_1$, $b_1$, and $c_1$) and three outer
principal semi-axes ($a_2$, $b_2$, and $c_2$) of the ellipsoidal
cloak  if there are no special instructions. We should also remark
that our simulation results are independent of the length scale of
the electric cloaks.


The streamlines of Fig.~\ref{fig1} represent the pathway of the
electric displacement field in the electric spherical cloak for the
parameters as indicated in the caption. Clearly, the displacement
field goes around the inner domain, but is limited within the cloak
(shell), and eventually returns to its original pathway. In this
process, an object inside the inner domain has no effect on the
displacement field at all. In other words, the object is protected
from the invasion of electric fields by using the electric cloak.
(Its cloaking efficiency will further be investigated in Section
3.2.)

For a spheroidal cloak, the pathway of the electric displacement
field is also illustrated in Fig.~\ref{fig2} (oblate spheroid) and
Fig.~\ref{fig3} (prolate spheroid). For the oblate spheroidal
cloak $a<b=c$, the pathway is similar to the spherical cloak
discussed in Fig.~\ref{fig1} when the applied electric field is
directed along the short principal semi-axis $a$, see
Fig.~\ref{fig2}. However, the situation becomes different for the
case of prolate spheroid with $a>b=c$, as shown in Fig.~\ref{fig3}
in which the external electric field is along the long principal
semi-axis $a$. Figure~\ref{fig3} shows a NGP (negative pathway)
 of the electric displacement field inside the spheroidal cloak. Here, the so-called NGP
means that the displacement field goes backwards, in contrast to
NMP (normal pathway) for which the field goes forwards. (All the
NGP streamlines shown in Fig.~\ref{fig3} are seemingly limited to
enter the inner domain. However, this is actually not true.
Depending on different streamlines/pathways and/or shape
parameters, such NGP streamlines can also appear within the cloak
domain only.) Owing to the symmetry of the prolate spheroid, the
NGP zone is strictly symmetrical, which locates close to the two
opposite sides of the prolate cloak.
On the other hand, we have also investigated the case of a general
ellipsoid with $a\neq b\neq c$ for which the external electric
field is directed along the longest principal semi-axis, the
behavior is generally similar to Fig.~\ref{fig3} (no figures shown
here). Nevertheless, in this case, the NGP zone is no longer
strictly symmetrical, the degree of which depends on the actual
values of $a$, $b$, and $c$.

Now let us dig out the physical origin of the NGP mentioned above.
It is known that an electric displacement field ${\bf D}$ is given
by ${\bf D} = \epsilon_0 {\bf E}+ {\bf P},$ where ${\bf E}$ is the
external electric field, and $P$ an electric polarization.
Corresponding to the NGP, ${\bf D}$ has a negative sign, which is
opposite to ${\bf E}$ (Strictly, there is an obtuse angle between
${\bf D}$ and ${\bf E}$ in the present case, due to the presence
of anisotropic and inhomogeneous electric permittivities inside
the cloak). Thus, this requires that ${\bf P}$ should be
sufficiently negative, satisfying ${\bf P} < -\epsilon_0 {\bf E}$.
So far, we can conclude that the appearance of the NGP zone
results from the shape-enhanced strong negative electric
polarization.


Figure~\ref{fig5}(a) shows  a phase diagram for spheroidal cloak
with three principal semi-axes $a\neq b=c$, in the presence of
external electric fields along the direction of the principal
semi-axis $a$. For this phase diagram, the potential difference
$U_{AB}$ (see the caption for its definition) is taken to be the
order parameter, and the ratio between $a$ and $b$ ($a: b$) the
tuning parameter ranging from 0 to 3.6. In the typical region of
$a: b$, a phase transition phenomenon comes to appear. In detail,
there is an existence of two distinct phases, NMP phase and NGP
phase, and a transition region between them, in which  the NGP is
enhanced significantly. Such a phase transition is driven by
geometrical shape of the cloak. To understand more clearly, in
Fig.~\ref{fig5}(b) we investigate $U_{AB}$ as a function of $a: b$
ranging from 0 to 20. It is evident that when $a: b$ is smaller
than or equal to 1 (oblate spheroid or sphere), the electric
potential $U_{AB}=0$, namely, there is no NGP. As $a: b>1$
(prolate spheroid), the NGP comes to appear, and reach maximum at
about $a: b=5: 2$. Eventually, it tends to disappear as $a: b$ is
large enough corresponding to a rod-like shape.


Figure~\ref{fig6} illustrates the paired electric permittivities,
$\epsilon_{\lambda}$ and $\epsilon_{\mu}$, in the NGP streamline
within the cross section of the spheroidal cloak with  $a> b=c$ as
a function of the distance $r$ (see the caption), for $a: b=$ (a)
$2: 1$, (b) $3: 1$, and (c) $4: 1$. All the
 points displayed in (a)-(c) have been taken in sequence from
the starting point to the ending point of the NGP streamline, as
indicated by the  ordered numbers. As $a: b$ is given, all the
paired $\varepsilon_{\lambda}$ and $\varepsilon_{\mu}$
(corresponding to the same number) appear in a symmetric
oscillation trajectory, and satisfy a sum rule: (a)
$\varepsilon_{\lambda} + \varepsilon_{\mu} = 2.015 \pm 0.003$, (b)
$2.218 \pm 0.007$, and (c) $2.461 \pm 0.006$. It is worth noting
that the electric permittivity $\varepsilon_{\nu}$ keeps unchanged
for the cross section (in two dimensions) of our interest.



3.2 Cloaking efficiency

We define the cloaking efficiency $\eta$ as the total electric
energy of domains II and III over the whole system of domains I, II,
and III. Apparently, if $\eta$ is equal to 100\%, the cloaking is
perfect. In this case, any objects within domain I can be perfectly
shielded from  external electric fields. In other words, the degree
of imperfectness of the cloaking is denoted by how $\eta$ deviates
from $100\% .$
For consistence, the applied electric potentials and the electric
permittivities of domains I, II, and III  are  set to be the same as
those already presented in Section 3.1.
In the simulations, the volume ratio between domain II and domain
I is also kept at 7:1.

The calculated cloaking efficiencies of the electric cloaks under
different conditions are shown in Tables I-III. As we can see, the
cloaking efficiency of the spherical cloak is 99.58\%, which is
very close to, but not equal to 100\% (perfect cloaking). The
deviation from 100\% arises from the small amount of central
streamlines that unavoidably pass through domain I. Further, we
find that the cloaking efficiencies of non-spherical cloaks are
always smaller than that of the spherical cloak. In detail, for
prolate spheroidal ($a>b=c$), oblate spheroidal ($a<b=c$), and
ellipsoidal ($a\neq b \neq c$) cloaks, their cloaking efficiencies
corresponding to three different principal semi-axes have only
slight difference and are smaller than that of the spherical
cloak. For the three kinds of non-spherical cloaks, we can
conclude that the cloaking efficiency corresponding to the field
directed along the longest principal semi-axes is the smallest
among those along the three principal semi-axes, and that  the
cloaking efficiency of ellipsoidal ($a\neq b \neq c$) cloaks is
the largest whereas that of oblate spheroidal ($a<b=c$) cloaks is
the smallest.


4 Discussion and conclusion

Besides DC electric fields adopted in the above simulations, our
results  also hold for AC fields due to the symmetry of the
system. Since they have been obtained by using numerical
simulations according to a first-principles approach, namely, the
coordinate transformation method, they are exact without any
approximation (despite the error which might occur when we
extracted the  data, e.g., as implied by the sum rules obtained
from Fig.~\ref{fig6}).
 Regarding
an experimental demonstration, it would be quite convenient to use
metamaterials to realize the electric cloaks with a shape-driven
phase transition between the NGP and NMP, due to the existing
significant achievements in the field~\cite{SmithS04,SchurigS06}.
Further, the fact that our results do not depend on specific
length-scales makes the experiment more tractable.




To sum up,   we have accurately revealed electric cloaks'
universality concerning a new class of shape-driven phase
transitions between the NGP and NMP of electric displacement
fields. We have showed that the NGP results from shape-enhanced
strong negative electric polarization, and that it corresponds to
a symmetric oscillation of the paired electric permittivities,
which have been demonstrated to satisfy a sum rule. The cloaking
efficiency has also been calculated for various geometrical
shapes.




We thank W. J. Tian for fruitful discussion. This work was
supported by the Pujiang Talent Project (No. 06PJ14006) of the
Shanghai Science and Technology Committee, by the Shanghai
Education Committee and the Shanghai Education Development
Foundation (Shuguang project),   by the National Natural Science
Foundation of China under Grant No.~10604014, and by Chinese
National Key Basic Research Special Fund under Grant
No.~2006CB921706.

\clearpage
\newpage
Figure Captions

Fig.~1.  (color online) Spherical cloak (left panel) and its cross
section (right panel). The streamlines denote the pathway of the
electric displacement field when the applied electric potentials are
set to be +3\,V and -3\,V at the top and bottom plane of the cubic
box, respectively. Parameters: $r_1=0.1\,$m (inner radius) and
$r_2=0.2\,$m (outer radius).

Fig.~2.  (color online) Oblate spheroidal cloak with three principal
semi-axes $a$, $b$, and $c$ satisfying $a<b=c$ (left panel) and its
cross section (right panel). The streamlines denote the pathway of
the electric displacement field in the direction of the short axis
$a$ when the applied electric potentials  are set to be +3\,V and
-3\,V at the top and bottom plane of the cubic box. Parameters:
$a_1=0.1\,$m and $b_1=c_1=0.2\,$m (three inner principal semi-axes);
$a_2=0.2\,$m and $b_2=c_2=0.4\,$m (three outer principal semi-axes).

Fig.~3.  (color online) Same as Fig.~\ref{fig2}, but for prolate
spheroidal cloak $a>b=c$ and the electric displacement field in the
direction of the long axis $a$. Parameters: $a_1=0.2\,$m and
$b_1=c_1=0.1\,$m; $a_2=0.4\,$m and $b_2=c_2=0.2\,$m.


Fig.~4.  (color online) (a) Phase diagram for spheroidal cloak with
three principal semi-axes $a\neq b=c$, which represents the
existence of two distinct phases, NMP phase and NGP phase, and a
transition region between them. The potential difference $U_{AB}$
between the starting point $A$ and ending point $B$ of the NGP
streamline [as also indicated in the inset of (b)] is taken to be
the order parameter, and the ratio between $a$ and $b$ ($a:b$) the
tuning parameter ranging from 0 to 3.6. The three curves correspond
to the three incident electric fields, which respectively have a
vertical distance of $0.8 b$, $0.5 b$, and $0.3 b$ with respect to
the principal semi-axis $a$, see also the inset of (b). (b) $U_{AB}$
as a function of $a:b$ ranging from 0 to 20, and others are the same
as (a).

Fig.~5.  (color online) The paired electric permittivities,
$\epsilon_{\lambda}$ and $\epsilon_{\mu}$, in the NGP streamline
within the cross section of the spheroidal cloak with  $a> b=c$ as a
function of the distance $r$ between any point in the NGP streamline
and the center of the cloak, for (a) $a:b = 2:1$, (b) $a:b = 3:1$,
and (c) $a:b = 4:1$. All the 20 points have been taken in sequence
from the starting point to the ending point of the NGP streamline,
as clearly indicated in the corresponding ordered numbers, see also
the three insets.

\clearpage
\newpage

\clearpage
\newpage

\clearpage
\newpage

\clearpage
\newpage

\clearpage
\newpage

\clearpage
\newpage
\begin{table}
\caption{Cloaking efficiency of  spherical and prolate spheroidal
($a>b=c$) cloaks. The external DC electric field is applied along
the $a$, $b$, and  $c$ principal semi-axes, respectively. $r_1$
(or $r_2$) denotes the inner (or outer) radius of a spherical
cloak.}
\begin{tabular}{lllllll}
\hline\hline
& shape  ~~~~~~~direction& inner & outer &$\eta$ \\
\hline
& sphere  & $r_1=0.1\,$m  & $r_2=0.2\,$m & 99.58\%  \\
\hline
&spheroid  ~~~~~~~$a$ & $a_1=0.2\,$m, $b_1=0.1\,$m, $c_1=0.1\,$m & $a_2=0.4\,$m, $b_2=0.2\,$m, $c_2=0.2\,$m & 99.14\% \\
\hline
&spheroid  ~~~~~~~$b$ or $c$  ~~~~~~~~& $a_1=0.2\,$m, $b_1=0.1\,$m, $c_1=0.1\,$m  ~~~~~~~~& $a_2=0.4\,$m, $b_2=0.2\,$m, $c_2=0.2\,$m ~~~~~~~~ & 99.15\% \\
 \hline\hline
\end{tabular}\label{table}
\end{table}

\clearpage
\newpage

\begin{table}
\caption{Cloaking efficiency of oblate spheroidal ($a<b=c$) cloaks.
}
\begin{tabular}{lllllll}
\hline\hline
& shape  ~~~~~~~direction& inner & outer &$\eta$ \\
\hline
&spheroid  ~~~~~~~$a$ & $a_1=0.1\,$m, $b_1=0.2\,$m, $c_1=0.2\,$m & $a_2=0.2\,$m, $b_2=0.4\,$m, $c_2=0.4\,$m & 97.75\% \\
\hline
&spheroid  ~~~~~~~$b$ or $c$ & $a_1=0.1\,$m, $b_1=0.2\,$m, $c_1=0.2\,$m & $a_2=0.2\,$m, $b_2=0.4\,$m, $c_2=0.4\,$m & 97.69\% \\
 \hline\hline
\end{tabular}\label{table}
\end{table}

\clearpage
\newpage

\begin{table}
\caption{Cloaking efficiency of  ellipsoidal ($a\neq b\neq c$)
cloaks. }
\begin{tabular}{lllllll}
\hline\hline
& shape ~~~~~~~direction& inner & outer &$\eta$ \\
\hline
&ellipsoid  ~~~~~~~$a$ ~~~~~~~~ & $a_1=0.1\,$m, $b_1=0.2\,$m, $c_1=0.4\,$m  ~~~~~~~~& $a_2=0.2\,$m, $b_2=0.4\,$m, $c_2=0.8\,$m  ~~~~~~~~& 99.55\% \\
\hline
&ellipsoid  ~~~~~~~$b$ & $a_1=0.1\,$m, $b_1=0.2\,$m, $c_1=0.4\,$m & $a_2=0.2\,$m, $b_2=0.4\,$m, $c_2=0.8\,$m & 99.56\% \\
\hline
&ellipsoid  ~~~~~~~$c$ & $a_1=0.1\,$m, $b_1=0.2\,$m, $c_1=0.4\,$m & $a_2=0.2\,$m, $b_2=0.4\,$m, $c_2=0.8\,$m & 99.52\% \\
 \hline\hline
\end{tabular}\label{table}
\end{table}


\end{document}